\documentclass[useAMS,usenatbib,intlimits,centertags,usegraphicx]{mn2e}
\usepackage{amsmath,amssymb}
\usepackage[latin1]{inputenc}
\usepackage{epsfig}
\usepackage{booktabs}

\usepackage{lscape}
\usepackage{array}
\newcommand{\PBS}[1]{\let\temp=\\#1\let\\=\temp}

\renewcommand{\d}{\mathrm d}
\newcommand{\rs}{r_{\mathrm s}(\chi)}
\newcommand{\rss}{r_{\mathrm s}(\chi')}

\begin{document}

\title{Uniformly Rotating Homogeneous Rings in Newtonian Gravity}

\author[Horatschek \& Petroff]
 {Stefan Horatschek\thanks{E-mail: {\tt S.Horatschek@tpi.uni-jena.de} (SH);\newline {\tt D.Petroff@tpi.uni-jena.de} (DP)} and David Petroff\addtocounter{footnote}{-1}\footnotemark\\
  Theoretisch-Physikalisches Institut, University of Jena,
                Max-Wien-Platz 1, 07743 Jena, Germany}

\pagerange{\pageref{firstpage}--\pageref{lastpage}} \pubyear{2008}

\maketitle

\label{firstpage}

\begin{abstract}
In this paper, we describe an analytical method for treating uniformly rotating homogeneous rings
without a central body in Newtonian gravity. We employ series expansions about the thin ring limit and use the fact that in this limit the cross-section of the ring tends to a circle.
The coefficients can in principle be determined up to an arbitrary order.
Results are presented here to the 20th order and compared with numerical results.
\end{abstract}


\section{Introduction}
The problem of the self-gravitating ring captured the interest of such renowned scientists as
\citet{Kowalewsky85}, \citet{Poincare85} and \citet{Dyson92, Dyson93}. Each of them tackled
the problem of an axially symmetric, homogeneous ring in equilibrium by expanding it about
the thin ring limit. In particular, Dyson provided a solution to fourth order in the
parameter $\sigma=a/b$, where $a$ provides a measure for the radius of the cross-section
of the ring and $b$ the distance of the cross-section's centre of mass from the axis of rotation.
\par
First numerical results were given by \citet{Wong74}, who was not able to clarify the
transition to spheroidal bodies. \citet{ES81} and \citet{EH85} developed improved methods
with which they were able to study this transition and the connection
to the Maclaurin spheroids. Returning to the problem significantly later, \citet*{AKM03} achieved
near-machine accuracy, which allowed them to study bifurcation sequences in detail and correct
erroneous results.
\par
In this paper we provide a scheme to extend Dyson's work up to arbitrary order in $\sigma$.
With the help of computer algebra\footnote{We made use of Maple\texttrademark. Maple is a
trademark of Waterloo Maple Inc.}, we were able to compute the solution explicitly up to
the 20th order and see a marked improvement over the fourth order. Although constant mass density
is of considerable importance to the method presented here, an extension of it to other
equations of state is possible \citep{PH08}.

\section{The Approximation Scheme}

\subsection{The Coordinates}
\begin{figure}
\centerline{\epsfig{file=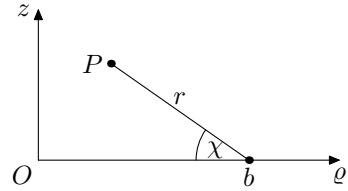}}
\caption{\label{Abb_coord} A sketch providing the meaning of the coordinates $(r,\chi)$ in relation
 to the cylindrical coordinates $(\varrho,z)$.}
\end{figure}

To describe axially symmetric rings, we introduce the polar-like coordinates $(r,\chi,\varphi)$,
which are related to the cylindrical coordinates $(\varrho,z,\varphi)$ by
\begin{align}
\varrho=b-r\cos\chi,\quad z=r\sin\chi,\quad \varphi=\varphi,
\end{align}
see also Fig.~\ref{Abb_coord}.
The surface of the ring will be described by a function $\rs$.
In our coordinates, the Laplace operator applied to a function $f=f(r,\chi)$ reads
\begin{align}
\begin{split}
\nabla^2f=&\frac{\partial^2f}{\partial r^2}+\frac1r\frac{\partial f}{\partial r}+\frac1{r^2}\frac{\partial^2f}{\partial\chi^2}\\
          &-(b-r\cos\chi)^{-1}\left(\cos\chi\frac{\partial f}{\partial r}-
            \frac{\sin\chi}r\frac{\partial f}{\partial\chi}\right).
\end{split}
\end{align}
We choose the constant $b$ in such a manner, that the centre of mass of the ring's cross-section
coincide with $r=0$, thus implying
\begin{align}
\label{com0}
\int_0^{2\pi}\int_0^{\rs}\mu r^2\cos\chi\,\d r\,\d\chi=0,
\end{align}
where $\mu$ is the mass density.

\subsection{Basic Equations}
For solving the problem of a self-gravitating fluid in equilibrium, we have to fulfil Laplace's equation
\begin{align}
\label{Lap}
\nabla^2 U=0
\end{align}
outside the fluid and Poisson's equation
\begin{align}
\label{Poi}
\nabla^2 U=4\pi G\mu
\end{align}
inside it, where $U$ is the gravitational potential.
Additionally, we have to satisfy Euler's equation
\begin{align}
\label{Eul}
\mu\frac{\d\bmath v}{\d t}=-\mu\nabla U-\nabla p,
\end{align}
where $\bmath v$ is the velocity of a fluid element and $p$ the pressure.
We consider uniform rotation about the axis $\varrho=0$ with the angular velocity $\mathbf\Omega$
and thus have the velocity field
\begin{align}
\bmath v=\mathbf\Omega\times\bmath x
\end{align}
leading to
\begin{align}
\label{eqi0}
\nabla\left(U+\int_0^p\frac{\d p'}{\mu(p')}-\frac12\Omega^2\varrho^2\right)=0.
\end{align}
Integration gives
\begin{align}
\label{eqi1}
U+\int_0^p\frac{\d p'}{\mu(p')}-\frac12\Omega^2\varrho^2=V_0,
\end{align}
where $V_0$ is the constant of integration.
At the surface of the ring, where the pressure vanishes, we have
\begin{align}
\label{eqi2}
U_\text{s}-\left.\frac12\Omega^2\varrho^2\right|_\text{s}=V_0.
\end{align}
We are dealing here with homogeneous rings, i.e.\ the mass density is a constant
\begin{align}
\label{EOS}
\mu=\text{constant}.
\end{align}
This means that the integral in \eqref{eqi0} and \eqref{eqi1} is simply $p/\mu$
and \eqref{com0} reduces to
\begin{align}
\label{com}
\int_0^{2\pi}\rs^3\cos\chi\,\d\chi=0.
\end{align}

\subsection{The Idea of the Approximation Scheme}
The thin ring limit is approached when the ratio
of the inner radius $\varrho_\text{i}$ to the outer one $\varrho_\text{o}$ tends to 1.
In this limit, the cross-section of the ring becomes a circle.
This is the starting point of the approximation.
As an interesting aside, if one considers a ring surrounding a central body, for example a point mass,
the cross-section of the ring can deviate significantly from a circle even if the radius ratio is close to 1,
see Fig.~8 in \citet{AP05}.
\par
We describe the surface of the ring by the Fourier series
\begin{align}
\rs=a\left(1+\sum_{i=1}^q\sum_{k=1}^i\beta_{ik}\cos(k\chi)\sigma^i+o(\sigma^q)\right),
\end{align}
where
\begin{align}
\sigma:=\frac{a}b.
\end{align}
There are no sine terms because of reflectional symmetry with respect to the
equatorial plane, which is known to hold for fluids in equilibrium
\citep[see][]{Lichtenstein33}. To the leading order, the cross-section
is indeed a circle of radius $a$.
We make a similar ansatz for the square of the angular velocity $\Omega$
\begin{align}
\Omega^2=\pi G\mu\left(\sum_{i=0}^{q+2}\Omega_i\sigma^i+o(\sigma^{q+2})\right)
\end{align}
and $V_0$
\begin{align}
\label{V0_ansatz}
V_0=-\pi G\mu a^2\left(\sum_{i=0}^q v_i\sigma^i+o(\sigma^q)\right).
\end{align}
\par
Because we consider constant mass density, a given surface function $\rs$
determines the distribution of the mass completely, i.e.\
one can compute the potential everywhere. For technical reasons,
we first calculate the potential on the axis of symmetry only.
After this, we determine the potential outside the ring and
in particular along its surface, where \eqref{eqi2} must hold.
\par
Since the potential outside the ring contains logarithmic terms in $r$ it will come as no surprise
that there are $\ln\sigma$ terms in the coefficients of our series. Like Dyson, we
introduce
\begin{align}
\lambda:=\ln\frac{8}\sigma-2.
\end{align}
\par
Let us suppose that we already know the solution up to the
$(q-1)$st order, and want to find the $q$th order, in particular
$\beta_{qk}$ ($k=1,2,\ldots,q$) and $\Omega_{q+1}$. To find these
$q+1$ unknowns, we have to fulfil equation \eqref{com} and also require that the
$q$ Fourier coefficients in front of $\cos(k\chi)$ ($k=1,2,\ldots,q$)
in \eqref{eqi2} vanish. After finding these $q+1$
unknowns, we now know $v_{q-1}$, but cannot yet determine the coefficient $v_q$.

\subsection{The Potential Outside the Ring}
First off, the Poisson integral is used to find the value of the potential along the
axis of rotation. We label the coordinates for a point on the axis
$(R,\chi_R)\equiv(r,\chi)$, from which
\begin{align}
\label{axis}
b=R\cos(\chi_R)
\end{align}
follows. The axis potential is
 \begin{align}
  U&_\text{axis}(R)= -2\pi G\int_0^{2\pi}\int_0^{\rss}\!\!\!\frac{\mu\,(b-r\cos \chi')r}{\sqrt{R^2+r^2-2Rr\cos\psi}}\,\d r\,\d\chi'\nonumber \\
     &= -2\pi G\mu\int_0^{2\pi}\!\int_0^{\rss}\!\!\!(b-r\cos\chi') \sum_{l=0}^\infty \left(\frac{r}{R}\right)^{l+1}\!\!P_l(\cos \psi)\,
         \d r\,\d\chi'\nonumber \\
     &=:-2\pi^2 G\mu a\sum_{l=1}^{\infty}(a^2\sigma^{-1})^l\frac{(2l-1)!!}{2l-1}\frac{A_l}{R^{2l-1}}
 \end{align}
 where $\psi:=\chi'-\chi_R$.
 Please note that the expansion in terms of powers of $1/R$ indicated in the
 last line is not trivial, since there is an $R$-dependence hidden in the terms
 with $\psi$. Because of reflectional symmetry, there are only terms with odd powers in $1/R$.
 We expand $A_l$ with respect to $\sigma$
 \begin{align}
  A_l=\sum_{i=l-1}^q\alpha_{li}\sigma^i+o(\sigma^q).
 \end{align}
 Using $U_\text{axis}$, we can then find the potential
 anywhere in the vacuum region. To do so, we first introduce a set of axially symmetric
 solutions to Laplace's equation that vanish at infinity. We define
 \begin{align}
  I_1(\varrho,z) &:=\int_0^\pi\!\frac{\d\varphi}{\sqrt{b^2+\varrho^2+z^2-2b\varrho\cos\varphi}},
 \end{align}
 which is nothing other than a multiple of the potential of a circular line
 of mass with radius $b$, centred around the axis. In $(r,\chi,\varphi)$-coordinates it reads
\begin{align}
 I_1(r,\chi)=\frac{2K\!\left(\sqrt{\frac{4b^2-4br\cos\chi}{4b^2-4br\cos\chi+r^2}}\right)}
                  {\sqrt{4b^2-4br\cos\chi+r^2}},
 \end{align}
 where $K$ denotes the complete elliptic integral of the first kind,
\begin{align}
K(k):=\int_0^{\frac{\pi}2}\frac{\d\theta }{\sqrt{1-k^2\sin^2\theta}}.
\end{align}
 Because the difference
 of two of such solutions with different $b$'s also satisfies Laplace's equation, it is clear that
\begin{align}
 I_l(r,\chi):=\left(\!-\frac{1}{b}\frac{\d}{\d b}\!\right)^{l-1}\!I_1(r,\chi),
 \end{align}
 where
 \begin{align}
\frac{\d}{\d b}=\frac{\partial}{\partial b}+\cos\chi\frac{\partial}{\partial r}-\frac{\sin\chi}r\frac{\partial}{\partial\chi}
 \end{align}
 is also a solution of Laplace's equation.
 Next we note that along the axis we have
 \begin{align}
  I_l(R) = \frac{\pi(2l-1)!!}{(2l-1)R^{2l-1}}.
 \end{align}
 It then follows that the potential in the vacuum region is
 \begin{align}
  \label{Uout}
  U_\text{out}(r,\chi)=-2\pi G\mu a^2\left(\sum_{l=1}^{q+1}a^{2l-1}\sigma^{-l}
  A_lI_l(r,\chi)+o(\sigma^q)\right),
 \end{align}
 since this expression satisfies Laplace's equation, vanishes at infinity and
 has the correct value along the axis. For calculating the potential at the body's surface
 we expand $I_l(r,\chi)$ for $r<b$. For example we get
 \begin{align}
  \label{I1}
   I_1(r,\chi)=\frac1b\left[\ln\left(\frac{8b}r\right)+
               \frac{[\ln(8b/r)-1]\cos\chi}{2}\frac{r}b+o\left(\frac{r}b\right)\right].
 \end{align}
 After evaluating these equations for $I_l(r,\chi)$ at the surface
 $r=\rs$, we use \eqref{Uout} to find the coefficients $\phi_{ik}$ in the expansion
 \begin{align}
 \label{Us}
 U_\text{s}(\chi)=-2\pi G\mu a^2\left(\sum_{i=0}^q\sum_{k=0}^i\phi_{ik}\cos(k\chi)\sigma^i
                    +o(\sigma^q)\right).
 \end{align}

\subsection{The Potential Inside the Ring and the Pressure}
To treat the interior of the ring, it is convenient to introduce
the dimensionless coordinate
\begin{align}
y:=\frac{r}a.
\end{align}
For calculating the potential inside the ring, we first expand it,
\begin{align}
\label{Uin}
U_\text{in}(r,\chi)=-\pi G\mu a^2\sum_{i=0}^q\sum_{k=0}^iU_{ik}(y)\cos(k\chi)\sigma^i.
\end{align}
Inserting this into Poisson's equation \eqref{Poi} and collecting the coefficients in front of
$\cos(k\chi)\sigma^i$ gives ordinary differential equations (ODEs) in $y$ for the $U_{ik}(y)$.
The solutions to these equations are determined uniquely by requiring that $U_{ik}(y)$ be finite
at the centre $y=0$ and that the potential be continuous at the surface,
i.e.\ $U_\text{in}(\rs,\chi)$ must agree with $U_\text{s}(\chi)$.
\par
After solving the ODEs for $U_{ik}(y)$, we know the inner potential, and
with \eqref{eqi1} we are able to calculate the pressure
\begin{align}
\label{p}
p(r,\chi)=\pi G\mu^2a^2\left(\sum_{i=0}^{q}\sum_{k=0}^ip_{ik}(y)\cos(k\chi)\sigma^i+o(\sigma^{q-1})\right).
\end{align}

\section{Solution up to the First Order}
To this order, we shall show the calculation in detail.
The unknown quantities are $\beta_{11}$, $\Omega_0$, $\Omega_1$, $\Omega_2$ and $v_0$.
The surface function reads
\begin{align}
\rs=a[1+\beta_{11}\sigma\cos\chi+o(\sigma)]
\end{align}
and equation \eqref{com} becomes
\begin{align}
\int_0^{2\pi}\rs^3\!\cos\chi\,\d\chi
=a^3[3\pi\beta_{11}\sigma+o(\sigma)].
\end{align}
This gives
\begin{align}
\beta_{11}=0,
\end{align}
cf.\ Table~\ref{tab_beta},
which means that we have the surface function
\begin{align}
\rs=a[1+o(\sigma)].
\label{rs_q1}
\end{align}
Next we have to evaluate equation \eqref{eqi2}. We immediately see that
\begin{align}
 \Omega_0=\Omega_1=0
\end{align}
must hold, and furthermore we have
\begin{align}
\label{v_squar}
\left.\frac12\Omega^2\varrho^2\right|_\text{s}=\pi G\mu a^2\left[\frac{\Omega_2}2+\left(\frac{\Omega_3}2-\Omega_2\cos\chi\right)\sigma+o(\sigma)
\right]
\end{align}
at the surface.
To calculate the potential at the ring's surface to the same order in $\sigma$, we first
compute the potential along the axis. Here we have to calculate the potential of a
torus\footnote{$E$ denotes the complete elliptic integral of the second kind,
$E(k):=\int_0^{\frac{\pi}2}\sqrt{1-k^2\sin^2\theta}\,\d\theta$.}:
\begin{align}
\begin{split}
U_\text{axis}(R)=-\frac{8\pi G\mu a^3}{3R}\bigg[\left(1+\frac{R^2}{a^2}\right)E\left(\frac{a}{R}\right)\sigma^{-1}\\
+\left(1-\frac{R^2}{a^2}\right)K\left(\frac{a}{R}\right)\sigma^{-1}+o(1)\bigg].
\end{split}
\end{align}
The expansion in terms of powers of $1/R$ gives
\begin{align}
A_1&=1+o(\sigma),\intertext{and}
A_2&=-\frac18\sigma+o(\sigma),
\end{align}
thus
\begin{align}
\alpha_{10}=1,\quad \alpha_{11}=0\quad\text{and}\quad\alpha_{21}=-\frac18,
\end{align}
cf.\ Table~\ref{tab_alpha}.
Now we are able to calculate the potential at the body's surface via \eqref{Uout},
\eqref{I1} and the corresponding equation for $I_2$. Using equation \eqref{rs_q1}
it is possible to expand the surface potential in $\sigma$. We get
\begin{align}
\label{Us_q1}
U_\text{s}(\chi)=-2\pi G\mu a^2\left[\lambda+2+\left(\frac{\lambda}2+\frac{3}8\right)\sigma\cos\chi+o(\sigma)\right],
\end{align}
which implies
\begin{align}
\phi_{00}=\lambda+2,\quad
\phi_{10}=0\quad\text{and}\quad
\phi_{11}=\frac{\lambda}{2}+\frac{3}8.
\end{align}
Plugging \eqref{V0_ansatz}, \eqref{v_squar} and \eqref{Us_q1} into equation \eqref{eqi2}
and collecting the coefficients in $\sigma^i\cos(k\chi)$, we find the following equations:
\begin{center}
 \renewcommand{\arraystretch}{1.2}
\begin{tabular}{ccl}
$i$ & $k$ & equation\\ \hline
0 & 0 & $2\lambda+4+\tfrac12\Omega_2=v_0$ \\
1 & 0 & $\tfrac12\Omega_3=v_1$ \\
1 & 1 & $\lambda+\tfrac34-\Omega_2=0$
\end{tabular}
\end{center}
Solving these equations gives
\begin{align}
\Omega_2&=\lambda+\tfrac34,
\intertext{cf.\ Table~\ref{tab_omega}, and}
v_0&=\tfrac52\lambda+\tfrac{35}8.
\end{align}
The equation with $\Omega_3$ and $v_1$ cannot be further evaluated until the next order $q=2$.
\par
For the mass and the angular momentum, we get
\begin{align}
\label{M}
M&=2\pi^2\mu a^3[\sigma^{-1}+o(\sigma^{-1})]\intertext{and}
J&=\sqrt{\pi^5G\mu^3\left(4\lambda+3\right)}\,{a}^{5}[\sigma^{-2}+o(\sigma^{-2})].
\end{align}
To leading order, $\varrho_\text{i}/\varrho_\text{o}=1-2\sigma$ holds,
and we can conclude that
\begin{align}
\lim_{\varrho_\text{i}/\varrho_\text{o}\to 1}\left[\frac{4\pi bV_0}{5GM}-\ln\left(1-\frac{\varrho_\text{i}}{\varrho_\text{o}}\right)\right]=\frac14-\ln 16,
\end{align}
see also equation (11) in \citet*{FHA05}.
A comparison of this result for homogeneous rings with the analogue for polytropic rings
can be found in \citet{PH08}.
\par
To calculate the inner potential, we
have to find a solution to Poisson's equation. The ansatz \eqref{Uin} leads to
the ODE
\begin{align}
\frac{\d^2U_{00}(y)}{\d y^2}+\frac1y\frac{\d U_{00}(y)}{\d y}+4=0
\end{align}
to leading order, which has the solution
\begin{align}
U_{00}(y)=-y^2+C_1\ln y+C_2.
\end{align}
At the centre, the potential has to be regular, thus $C_1=0$.
The resulting potential at the surface $r=\rs$ is
\begin{align}\label{Us_00}
U_\text{s}(\chi)=\pi G\mu a^2[1-C_2+o(1)].
\end{align}
For the potential to be continuous, $C_2=2\lambda+5$ must hold, see \eqref{Us_q1}.
To the zeroth order the potential is not a function of the angle $\chi$.
\par
Furthermore we have the equations
\begin{align}
\frac{\d^2U_{10}(y)}{\d y^2}+\frac1y\frac{\d U_{10}(y)}{\d y}=0
\end{align}
and
\begin{align}
\frac{\d^2U_{11}(y)}{\d y^2}+\frac1y\frac{\d U_{11}(y)}{\d y}-\frac{U_{11}(y)}{y^2}+2y=0.
\end{align}
Note that the $2y$ term in the second equation results from $U_{00}(y)$, which is already known.
The solutions of these equations that are regular at the centre and have the correct values at the surface are
\begin{align}
U_{10}(y)&=0\intertext{and}
U_{11}(y)&=(\lambda+1)y-\frac{y^3}4,
\end{align}
cf.\ Table~\ref{tab_Uinnen}.
For the pressure, \eqref{eqi1} leads to
\begin{align}
p_{00}(y)&=1-y^2,\\
p_{10}(y)&=0\intertext{and}
p_{11}(y)&=\frac{y}4-\frac{y^3}4.
\end{align}
After finding the inner potential and pressure, we can calculate the potential energy
\begin{align*}
W:=\frac{\mu}2\int U\,\d V
=-\pi^3G\mu^2a^5\left[\left(2\lambda+\frac{9}{2}\right)\sigma^{-1}+o(\sigma^{-1})\right],
\end{align*}
the rotational energy
\begin{align*}
T:=\frac{\mu\Omega^2}2\int\varrho^2\,\d V
=\pi^3G\mu^2a^5\left[\left(\lambda+\frac34\right)\sigma^{-1}+o(\sigma^{-1})\right]
\end{align*}
and the integral over the pressure
\begin{align}
P:=\int p\,\d V
=\pi^3G\mu^2a^5\left(\sigma^{-1}+o(\sigma^{-1})\right)
\end{align}
to first order.
We see that the virial theorem \eqref{virial} is fulfilled up to this order.

\begin{table*}
  \centering
  \begin{minipage}{140mm}
  \caption{For a given radius ratio $\varrho_\text i/\varrho_\text o=0.9$,
physical quantities to different orders in $q$ and numerically determined values
           are compared to the
           values for $q=20$:
$\bar M_{20}=4.6299179884304816293\times 10^{-2}$,
$\bar\Omega^2_{20}=3.2474683264953211610\times 10^{-2}$,
$\bar J_{20}=7.5456215256289320669\times 10^{-3}$,
$\bar P_{20}=1.7862946528142761708\times 10^{-4}$,
$\bar T_{20}=6.7988816964653749490\times 10^{-4}$,
$\bar W_{20}=-1.8956647351373578410\times 10^{-3}$.
  \label{tab09}}
  \begin{tabular}{cccccccc}\toprule
$q$ & $\sigma$ &  $\bar{M}_q/\bar{M}_{20}-1$ & $\bar{\Omega}^2_q/\bar{\Omega}^2_{20}-1$ &
$\bar{J}_q/\bar{J}_{20}-1$ & $\bar{P}_q/\bar{P}_{20}-1$ & $\bar{T}_q/\bar{T}_{20}-1$ & $\bar{W}_q/\bar{W}_{20}-1$\\ \midrule
$ 1$ & $0.053$ & $\phantom{+}1\times 10^{-2}$ & $\phantom{+}1\times 10^{-2}$ & $\phantom{+}2\times 10^{-2}$ & $\phantom{+}3\times 10^{-2}$ & $\phantom{+}2\times 10^{-2}$ & $\phantom{+}2\times 10^{-2}$ \\
$ 2$ & $0.052$ & $\phantom{+}2\times 10^{-4}$ & $\phantom{+}6\times 10^{-4}$ & $\phantom{+}4\times 10^{-4}$ & $\phantom{+}6\times 10^{-3}$ & $-1\times 10^{-3}$ & $\phantom{+}6\times 10^{-4}$ \\
$ 3$ & $0.052$ & $\phantom{+}2\times 10^{-4}$ & $\phantom{+}1\times 10^{-4}$ & $\phantom{+}2\times 10^{-4}$ & $\phantom{+}4\times 10^{-4}$ & $\phantom{+}3\times 10^{-4}$ & $\phantom{+}3\times 10^{-4}$ \\
$ 4$ & $0.052$ & $\phantom{+}3\times 10^{-6}$ & $-1\times 10^{-5}$ & $-3\times 10^{-6}$ & $\phantom{+}5\times 10^{-5}$ & $-5\times 10^{-5}$ & $-2\times 10^{-5}$ \\
$10$ & $0.052$ & $\phantom{+}6\times 10^{-11}$ & $-9\times 10^{-11}$ & $\phantom{+}4\times 10^{-12}$ & $\phantom{+}5\times 10^{-10}$ & $-6\times 10^{-10}$ & $-3\times 10^{-10}$ \\
$19$ & $0.052$ & $\phantom{+}4\times 10^{-17}$ & $\phantom{+}4\times 10^{-17}$ & $\phantom{+}6\times 10^{-17}$ & $\phantom{+}1\times 10^{-16}$ & $\phantom{+}9\times 10^{-17}$ & $\phantom{+}9\times 10^{-17}$ \\
\midrule[0.2pt]
num& --- & $-4\times 10^{-14}$& $\phantom{+}1\times 10^{-16}$& $-4\times 10^{-14}$& $-7\times 10^{-14}$& $-4\times 10^{-14}$& $-3\times 10^{-14}$\\
\bottomrule
  \end{tabular}
  \end{minipage}
 \end{table*}

\begin{table*}
  \centering
  \begin{minipage}{140mm}
  \caption{For a given radius ratio $\varrho_\text i/\varrho_\text o=0.5$,
physical quantities to different orders in $q$
are compared to the numerically determined values
$\bar M_\text{num}=0.7201292$,
$\bar\Omega^2_\text{num}=0.5467604$,
$\bar J_\text{num}=0.3247949$,
$\bar P_\text{num}=0.04874713$,
$\bar T_\text{num}=0.1200820$,
$\bar W_\text{num}=-0.3864053$.
  \label{tab05}}
  \begin{tabular}{cccccccc}\toprule
$q$ & $\sigma$ & $\bar{M}_q/\bar{M}_\text{num}-1$ & $\bar{\Omega}^2_q/\bar{\Omega}^2_\text{num}-1$ &
$\bar{J}_q/\bar{J}_\text{num}-1$ & $\bar{P}_q/\bar{P}_\text{num}-1$ & $\bar{T}_q/\bar{T}_\text{num}-1$ & $\bar{W}_q/\bar{W}_\text{num}-1$\\ \midrule
$ 1$ & $0.33$ & $2.8\times 10^{-1}$ & $2.3\times 10^{-1}$ & $4.2\times 10^{-1}$ & $8.6\times 10^{-1}$ & $4.6\times 10^{-1}$ & $6.1\times 10^{-1}$ \\
$ 2$ & $0.30$ & $5.1\times 10^{-2}$ & $5.9\times 10^{-2}$ & $6.5\times 10^{-2}$ & $2.5\times 10^{-1}$ & $2.6\times 10^{-2}$ & $1.1\times 10^{-1}$ \\
$ 3$ & $0.30$ & $5.1\times 10^{-2}$ & $4.2\times 10^{-2}$ & $5.8\times 10^{-2}$ & $1.3\times 10^{-1}$ & $8.1\times 10^{-2}$ & $1.0\times 10^{-1}$ \\
$ 4$ & $0.30$ & $1.7\times 10^{-2}$ & $1.3\times 10^{-2}$ & $1.9\times 10^{-2}$ & $5.4\times 10^{-2}$ & $8.4\times 10^{-3}$ & $2.6\times 10^{-2}$ \\
$10$ & $0.29$ & $1.2\times 10^{-3}$ & $1.0\times 10^{-3}$ & $1.3\times 10^{-3}$ & $3.2\times 10^{-3}$ & $1.0\times 10^{-3}$ & $1.9\times 10^{-3}$ \\
$20$ & $0.29$ & $2.6\times 10^{-5}$ & $2.5\times 10^{-5}$ & $3.0\times 10^{-5}$ & $7.2\times 10^{-5}$ & $2.7\times 10^{-5}$ & $4.4\times 10^{-5}$ \\
\bottomrule
  \end{tabular}
  \end{minipage}
 \end{table*}

\begin{table*}
  \centering
  \begin{minipage}{140mm}
  \caption{For a given radius ratio $\varrho_\text i/\varrho_\text o=0.2$,
physical quantities to different orders in $q$
are compared to the numerically determined values
$\bar M_\text{num}=0.9424$,
$\bar\Omega^2_\text{num}=0.9844$,
$\bar J_\text{num}=0.4545$,
$\bar P_\text{num}=0.07865$,
$\bar T_\text{num}=0.2255$,
$\bar W_\text{num}=-0.6869$.
  \label{tab02}}
  \begin{tabular}{cccccccc}\toprule
$q$ & $\sigma$ & $\bar{M}_q/\bar{M}_\text{num}-1$ & $\bar{\Omega}^2_q/\bar{\Omega}^2_\text{num}-1$ &
$\bar{J}_q/\bar{J}_\text{num}-1$ & $\bar{P}_q/\bar{P}_\text{num}-1$ & $\bar{T}_q/\bar{T}_\text{num}-1$ & $\bar{W}_q/\bar{W}_\text{num}-1$\\ \midrule
$ 1$ & $0.67$ & $1.0$ & $7.5\times 10^{-1}$ & $1.6$ & $5.1$ & $1.6$ & $2.8$ \\
$ 2$ & $0.54$& $3.2\times 10^{-1}$ & $3.5\times 10^{-1}$ & $3.9\times 10^{-1}$ & $1.6$ & $3.2\times 10^{-1}$ & $7.6\times 10^{-1}$ \\
$ 3$ & $0.54$& $3.2\times 10^{-1}$ & $2.7\times 10^{-1}$ & $3.5\times 10^{-1}$ & $1.1$ & $5.4\times 10^{-1}$ & $7.2\times 10^{-1}$ \\
$ 4$ & $0.51$ & $1.9\times 10^{-1}$ & $1.7\times 10^{-1}$ & $2.1\times 10^{-1}$ & $6.3\times 10^{-1}$ & $2.1\times 10^{-1}$ & $3.6\times 10^{-1}$ \\
$10$ & $0.47$ & $6.4\times 10^{-2}$ & $6.9\times 10^{-2}$ & $7.1\times 10^{-2}$ & $1.9\times 10^{-1}$ & $8.7\times 10^{-2}$ & $1.2\times 10^{-1}$ \\
$20$ & $0.46$ & $2.2\times 10^{-2}$ & $2.7\times 10^{-2}$ & $2.3\times 10^{-2}$ & $6.0\times 10^{-2}$ & $3.2\times 10^{-2}$ & $4.2\times 10^{-2}$ \\
\bottomrule
  \end{tabular}
  \end{minipage}
 \end{table*}

\section{Discussion}
With this approximation method, we are able to calculate e.g.\ the shape, angular velocity
and pressure of the ring up to arbitrary order in $\sigma$. We have done so up to
the 20th order.
\par
To test our solutions, we ensured that the transition condition
\begin{align}
\nabla U_\text{in}|_\text{s}=\nabla U_\text{out}|_\text{s}
\end{align}
is fulfilled up to the appropriate order in $\sigma$.
Furthermore we tested that the virial theorem
\begin{align}
\label{virial}
W+2T+3P=0
\end{align}
is fulfilled for each order in $\sigma$.
\par
An important question is how good this method is.
In Tables~\ref{tab09}, \ref{tab05} and \ref{tab02} one can see how the dimensionless
quantities
\begin{align}
 \begin{split}
 \frac{\bar{M}}{M}=\frac{1}{\mu\varrho_\text{o}^3}, \qquad
 \frac{\bar{\Omega}^2}{\Omega^2}&=\frac{1}{G\mu},     \qquad
 \frac{\bar{J}}{J}=\frac{1}{G^{1/2}\mu^{3/2}\varrho_\text{o}^5},\\
 \frac{\bar{P}}{P}=\frac{\bar{T}}{T}&=\frac{\bar{W}}{W}=
         \frac{1}{G\mu^2\varrho_\text{o}^5}
 \end{split}
\end{align}
improve in accuracy with increasing order for different radius ratios.
Especially for thin rings, we get very accurate results. In fact, for rings with radius ratios
$\varrho_\text{i}/\varrho_\text{o}\approx 0.85$ we achieve a precision
which is comparable with that given by the numerical method described in \citet*{AKM03b}.
For larger radius ratios, the accuracy is thus better.
As a co-product, our work provides an independent test of the
accuracy of the numerical method (better than $10^{-13}$ cf.\ Table~\ref{tab09}).
\par
\begin{figure}
\centerline{\epsfig{file=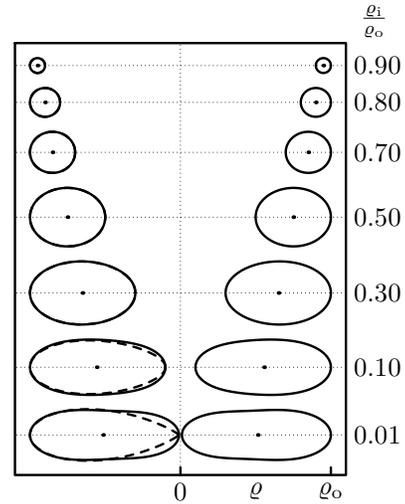}}
\caption{\label{Abb_cs} Meridional cross-sections of rings to the order $q=20$ for different radius ratios $\varrho_\text{i}/\varrho_\text{o}$. The $\varrho$- and $z$-axis are scaled identically in such a manner that $\varrho_\text{o}$ has the same value for all the rings. The dot in each ring marks the centre
of mass of the cross-section ($\varrho=b, z=0$ i.e. $r=0$) and the dashed line shows the numerical
result and is indistinguishable from the $q=20$ curve for
$\varrho_\text{i}/\varrho_\text{o}\ga 0.3$.}
\end{figure}
\begin{figure}
\centerline{\epsfig{file=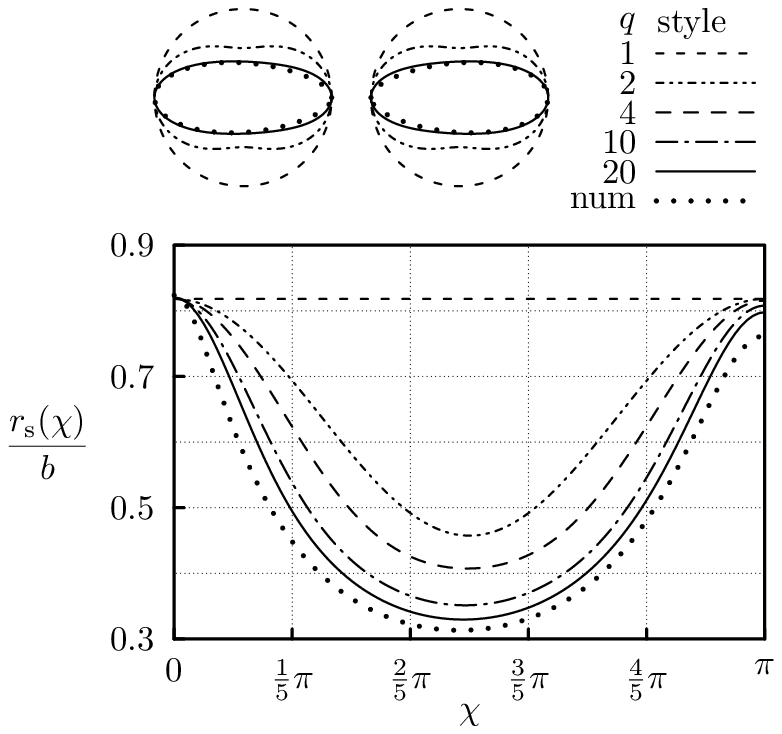}}
\caption{\label{Abb_cs2}
The meridional cross-section and the (dimensionless) surface function $\rs/b$
of the ring with radius ratio $\varrho_\text{i}/\varrho_\text{o}=0.1$ for different orders $q$ compared
to the numerical result. The surfaces are scaled such that $\varrho_\text{o}$ (and therefore also
$\varrho_\text{i}$) has the same value to all orders.}
\end{figure}
\begin{figure}
\centerline{\epsfig{file=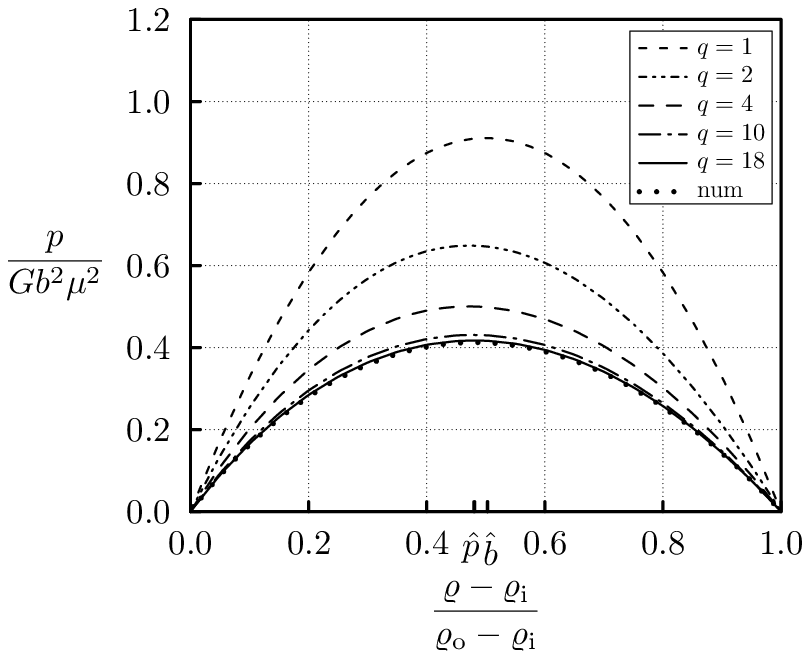}}
\caption{\label{Abb_p} The pressure in the equatorial plane for a ring with radius ratio
 $\varrho_\text{i}/\varrho_\text{o}=0.3$ for different orders $q$ compared to the numerical result.
 It is interesting to note that the centre of mass of the cross-section
 does not coincide with the point of maximum pressure.
 To the order $q=18$ we get $\hat b:=(b-\varrho_\text{i})/(\varrho_\text{o}-\varrho_\text{i})=0.503$
 and $\hat p:=(\varrho_\text{p,max}-\varrho_\text{i})/(\varrho_\text{o}-\varrho_\text{i})=0.480$, which
 differ by less than 1\% from the numerical values.}
\end{figure}
\begin{figure}
\centerline{\epsfig{file=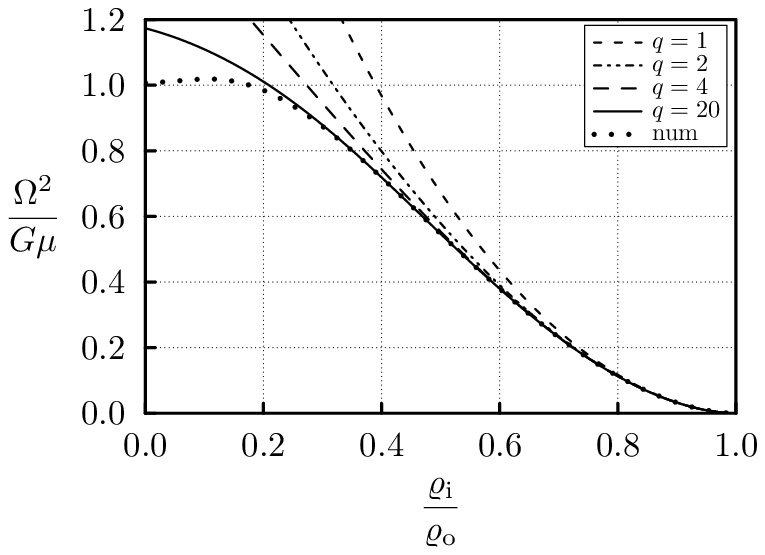}}
\caption{\label{Abb_om} The (dimensionless) squared angular velocity $\Omega^2/G\mu$ as function of the
 radius ratio $\varrho_\text{i}/\varrho_\text{o}$ for different orders $q$ compared to the numerical
 result.}
\end{figure}

The shape of the ring in meridional cross-section for various radius ratios can be found in
Fig.~\ref{Abb_cs}. The curves to order $q=20$ can barely be distinguished from the numerical
ones for $\varrho_\text i/\varrho_\text o\ga 0.3$. As one approaches the transition to spheroidal
topologies ($\varrho_\text i/\varrho_\text o\to 0$), the true curve becomes pointy at the inner edge and is no longer well represented by our Fourier series.
Nevertheless, the shape of the ring is quite well approximated even for $\varrho_\text{i}/\varrho_\text{o}=0.1$, as seen in Fig.~\ref{Abb_cs2}.
The surface function $\rs$, which is a constant to leading order, clearly
approaches the numerical one with increasing $q$.
The pressure in the equatorial plane can also be seen to approach the numerically
determined one for $\varrho_\text{i}/\varrho_\text{o}=0.3$ in Fig.~\ref{Abb_p}.
It is interesting to note that the centre of mass does not coincide with the
point of maximum pressure.
In Fig.~\ref{Abb_om} one can get an impression of the accuracy of the approximation over the whole
range of radius ratios and for various values of $q$.
Despite the claim found in \citet{Wong74} that Dyson's perturbative method
diverges for $\sigma>1/3$ \citep[see also the comments in][]{Dyson92},
these results indicate the opposite.
\par
The (dimensionless) coefficients $\beta_{ik}$ and $\Omega_i$ are only functions of the dimensionless
quantity $\sigma=a/b$, but not of any length like $\varrho_\text{o}$ for example. The reason for this
is that equations \eqref{Lap}, \eqref{Poi}, \eqref{Eul} and \eqref{EOS} are
scale invariant in the following sense: If $U(\bmath x)$, $\mu(\bmath x)$, $p(\bmath x)$ and
$\bmath v(\bmath x)$ are solutions of these equations then $\alpha^{-2}U(\alpha\bmath x)$,
$\mu(\alpha\bmath x)$, $\alpha^{-2}p(\alpha\bmath x)$
and $\alpha^{-1}\bmath v(\alpha\bmath x)$ are too, where $\alpha$ is an arbitrary scaling factor.
Other equations of state $\mu=\mu(p)$ will not satisfy this scaling invariance
in general. To parametrize a specific ring, including its dimension,
we need two parameters (for example $\sigma$ and $b$) as expected.
\par
Please note that one has to be careful in interpreting the results for the thin ring limit.
For example, one might think that the squared angular velocity vanishes like $\sigma^2\ln\sigma$.
This is true for the dimensionless quantity $\Omega^2/G\mu$, but need not be true for the
squared angular velocity itself. If we fix the `size' $b$ and the mass $M$ of the ring
in that limit, then the cross-section shrinks to a point ($a=\sigma b$).
With \eqref{M} we can conclude that $\mu\propto\sigma^{-2}$ and therefore
$\Omega^2\propto\ln\sigma$, which means that $\Omega^2$ and hence
the velocity of a fluid element go to infinity.
\par
Relativistic rings, including the thin ring limit, were studied in \citet*{AKM03c},
\citet{Ansorgetal04} and \citet{FHA05}.
From the perspective of General Relativity, the Newtonian theory constitutes
a good approximation when certain conditions are fulfilled. For one thing,
typical velocities must be small compared to the speed of light and
for another $G|U|\ll c^2$ must hold. We just saw, however, that for rings
of finite extent and mass, the velocities grow unboundedly in the thin
ring limit. The same holds for $U_\text s$ as well, see \eqref{Us_00}.
This means that the Newtonian theory of gravity is not
appropriate to describe this subtle limit itself, since one cannot expect
it to be a good approximation to General Relativity.
It is remarkable that the approximation about the point $\sigma=0$ is
nevertheless so successful.

\section*{Acknowledgments}
It is a pleasure to thank Reinhard Meinel for fruitful discussions.
This research was funded in part by the Deutsche Forschungsgemeinschaft
(SFB/TR7--B1).

 \bibliographystyle{mn2e}
 \bibliography{Reflink}

 \appendix
\section{Further coefficients}

In Tables~\ref{tab_omega}--\ref{tab_alpha}, coefficients for $\Omega_i$, $\beta_{ik}$, $U_{ik}(y)$ and $\alpha_{li}$ are given.

\newpage

\begin{table}
 \renewcommand{\arraystretch}{1.8}\renewcommand{\baselinestretch}{1.4}
 \centering
\caption{Coefficients $\Omega_i$ up to the order $q=9$ ($i=2,3,\ldots,10$).\label{tab_omega}}
  \begin{tabular}{c|*{1}{>{\footnotesize\PBS\raggedright}p{7cm}}}
$i$ & $\Omega_i$ \\\hline2 & $\lambda+\frac{3}{4}
$\\3 & $0
$\\4 & $-\frac{1}{8}\,\lambda-{\frac {19}{96}}
$\\5 & $0
$\\6 & ${\frac {25}{128}}\,{\lambda}^{3}+{\frac {365}{1536}}\,{\lambda}^{2}-{
\frac {2345}{18432}}\,\lambda-{\frac {8989}{73728}}
$\\7 & $0
$\\8 & ${\frac {25}{64}}\,{\lambda}^{4}+{\frac {40235}{98304}}\,{\lambda}^{3}-
{\frac {134255}{294912}}\,{\lambda}^{2}-{\frac {36493505}{56623104}}\,
\lambda-{\frac {34831813}{226492416}}
$\\9 & $0
$\\10 & ${\frac {134925}{131072}}\,{\lambda}^{5}+{\frac {2797535}{1572864}}\,{
\lambda}^{4}-{\frac {50072105}{169869312}}\,{\lambda}^{3}-{\frac {
1021727845}{509607936}}\,{\lambda}^{2}-{\frac {104581877693}{
97844723712}}\,\lambda-{\frac {49377918425}{391378894848}}
$\\\end{tabular}

 \end{table}

\begin{landscape}
\begin{table}
\renewcommand{\arraystretch}{1.8}\renewcommand{\baselinestretch}{1.4}
\caption{Coefficients $\beta_{ik}$ up to the order $q=9$. The bold-faced type indicates that these terms
are incorrect in \citet{Dyson92}.\label{tab_beta}}
 \begin{tabular}{c|*{9}{>{\footnotesize\PBS\raggedright}p{2.00cm}}}
{\Large $ _i\!\!\diagdown\!\!^k$} & 1 & 2 & 3 & 4 & 5 & 6 & 7 & 8 & 9  \\  \hline 
1 & $0
$&---&---&---&---&---&---&---&---\\2 & $0
$&$\frac{5}{8}\,\lambda+{\frac {35}{96}}
$&---&---&---&---&---&---&---\\3 & $0
$&$0
$&${\frac {5}{128}}\,\lambda-{\frac {35}{3072}}
$&---&---&---&---&---&---\\4 & $0
$&$\frac{5}{8}\,{\lambda}^{2}+\mathbf{{\frac {95}{128}}\,\lambda+{\frac {1145}{9216}}}
$&$0
$&${\frac {75}{256}}\,{\lambda}^{2}+\mathbf{{\frac {815}{2304}}\,\lambda+{\frac {
5089}{55296}}}
$&---&---&---&---&---\\5 & $-{\frac {25}{1024}}\,{\lambda}^{2}-{\frac {175}{24576}}\,\lambda+{
\frac {1225}{294912}}
$&$0
$&${\frac {15}{256}}\,{\lambda}^{2}+{\frac {4955}{147456}}\,\lambda-{
\frac {62141}{3538944}}
$&$0
$&${\frac {25}{512}}\,{\lambda}^{2}+{\frac {75}{4096}}\,\lambda-{\frac {
14455}{1179648}}
$&---&---&---&---\\6 & $0
$&${\frac {5185}{4096}}\,{\lambda}^{3}+{\frac {110515}{49152}}\,{\lambda}
^{2}+{\frac {853225}{884736}}\,\lambda+{\frac {34487}{1327104}}
$&$0
$&${\frac {75}{128}}\,{\lambda}^{3}+{\frac {116545}{110592}}\,{\lambda}^{
2}+{\frac {308395}{589824}}\,\lambda+{\frac {17892169}{318504960}}
$&$0
$&${\frac {625}{4096}}\,{\lambda}^{3}+{\frac {169625}{589824}}\,{\lambda}
^{2}+{\frac {1097461}{7077888}}\,\lambda+{\frac {7327349}{339738624}}
$&---&---&---\\7 & $-{\frac {125}{2048}}\,{\lambda}^{3}-{\frac {76325}{1179648}}\,{\lambda
}^{2}+{\frac {78175}{28311552}}\,\lambda+{\frac {1313095}{169869312}}
$&$0
$&${\frac {1325}{8192}}\,{\lambda}^{3}+{\frac {109915}{442368}}\,{\lambda
}^{2}+{\frac {863633}{18874368}}\,\lambda-{\frac {105833029}{
4076863488}}
$&$0
$&${\frac {125}{1024}}\,{\lambda}^{3}+{\frac {9815}{73728}}\,{\lambda}^{2
}-{\frac {472843}{56623104}}\,\lambda-{\frac {32985941}{1509949440}}
$&$0
$&${\frac {375}{8192}}\,{\lambda}^{3}+{\frac {4825}{98304}}\,{\lambda}^{2
}-{\frac {34745}{9437184}}\,\lambda-{\frac {268795}{28311552}}
$&---&---\\8 & $0
$&${\frac {11935}{4096}}\,{\lambda}^{4}+{\frac {15693265}{2359296}}\,{
\lambda}^{3}+{\frac {387327415}{84934656}}\,{\lambda}^{2}+{\frac {
1967050099}{2717908992}}\,\lambda-{\frac {21240691225}{195689447424}}
$&$0
$&${\frac {48175}{32768}}\,{\lambda}^{4}+{\frac {299613275}{84934656}}\,{
\lambda}^{3}+{\frac {86354441}{31850496}}\,{\lambda}^{2}+{\frac {
163195481003}{244611809280}}\,\lambda+{\frac {61946667647}{
14676708556800}}
$&$0
$&${\frac {1875}{4096}}\,{\lambda}^{4}+{\frac {1994875}{1769472}}\,{
\lambda}^{3}+{\frac {1544984219}{1698693120}}\,{\lambda}^{2}+{\frac {
26196760631}{101921587200}}\,\lambda+{\frac {523810097561}{
34245653299200}}
$&$0
$&${\frac {21875}{262144}}\,{\lambda}^{4}+{\frac {2069375}{9437184}}\,{
\lambda}^{3}+{\frac {7862435}{42467328}}\,{\lambda}^{2}+{\frac {
6142108445}{114152177664}}\,\lambda+{\frac {968695327}{342456532992}}
$&---\\9 & $-{\frac {26875}{131072}}\,{\lambda}^{4}-{\frac {10803575}{28311552}}\,
{\lambda}^{3}-{\frac {73205885}{452984832}}\,{\lambda}^{2}+{\frac {
245384393}{10871635968}}\,\lambda+{\frac {5148689159}{391378894848}}
$&$0
$&${\frac {58565}{131072}}\,{\lambda}^{4}+{\frac {317810795}{339738624}}
\,{\lambda}^{3}+{\frac {4061560685}{8153726976}}\,{\lambda}^{2}-{
\frac {3261000013}{97844723712}}\,\lambda-{\frac {1033257753043}{
23482733690880}}
$&$0
$&${\frac {3125}{8192}}\,{\lambda}^{4}+{\frac {20750915}{28311552}}\,{
\lambda}^{3}+{\frac {738672683}{2264924160}}\,{\lambda}^{2}-{\frac {
180456831247}{3261490790400}}\,\lambda-{\frac {13297677452861}{
365286968524800}}
$&$0
$&${\frac {2625}{16384}}\,{\lambda}^{4}+{\frac {2597725}{9437184}}\,{
\lambda}^{3}+{\frac {112005563}{1358954496}}\,{\lambda}^{2}-{\frac {
117378627479}{2283043553280}}\,\lambda-{\frac {173703027451}{
9132174213120}}
$&$0
$&${\frac {625}{16384}}\,{\lambda}^{4}+{\frac {72125}{1048576}}\,{\lambda
}^{3}+{\frac {4955105}{226492416}}\,{\lambda}^{2}-{\frac {79046365}{
5435817984}}\,\lambda-{\frac {1636799939}{260919263232}}
$\\\end{tabular}

 \end{table}
\end{landscape}

 \begin{landscape}
 \begin{table}
 \renewcommand{\arraystretch}{1.8}\renewcommand{\baselinestretch}{1.4}
 \caption{Coefficients $U_{ik}(y)$ up to the order $q=7$.\label{tab_Uinnen}}
  \begin{tabular}{c|*{1}{>{\footnotesize\PBS\raggedright}p{3.31cm}}
                  *{1}{>{\footnotesize\PBS\raggedright}p{3.20cm}}
                  *{1}{>{\footnotesize\PBS\raggedright}p{2.78cm}}
                  *{1}{>{\footnotesize\PBS\raggedright}p{2.94cm}}
                  *{1}{>{\footnotesize\PBS\raggedright}p{2.31cm}}
                  *{1}{>{\footnotesize\PBS\raggedright}p{1.83cm}}
                  *{1}{>{\footnotesize\PBS\raggedright}p{1.57cm}}
                  *{1}{>{\footnotesize\PBS\raggedright}p{1.57cm}}
}
{\Large $ _i\!\!\diagdown\!\!^k$} & 0 & 1 & 2 & 3 & 4 & 5 & 6 & 7  \\  \hline
0 & $2\,\lambda +5 -{y}^{2}
$&---&---&---&---&---&---&---\\
1 & $0
$&$\big( \lambda+1 \big)y -\frac{1}{4}\,{y}^{3}
$&---&---&---&---&---&---\\
2 & $-\frac{1}{8}\,\lambda-{\frac {7}{32}}+ \big( \frac{1}{4}\,\lambda+\frac{1}{4} \big) {y}^{2}
-{\frac {3}{32}}\,{y}^{4}
$&$0
$&$ \big( \lambda+{\frac {31}{48}} \big){y}^{2} -{\frac {5}{48}}\,{y}^
{4}
$&---&---&---&---&---\\
3 & $0
$&$ \big( -{\frac {13}{32}}\,\lambda-{\frac {65}{192}} \big)y +
 \big( {\frac {7}{16}}\,\lambda+{\frac {67}{192}} \big) {y}^{3}-{
\frac {15}{128}}\,{y}^{5}
$&$0
$&$\big( {\frac {15}{64}}\,\lambda+{\frac {175}{1536}} \big){y}^{3} -
{\frac {35}{768}}\,{y}^{5}
$&---&---&---&---\\
4 & ${\frac {25}{64}}\,{\lambda}^{3}+{\frac {295}{384}}\,{\lambda}^{2}+{
\frac {3457}{9216}}\,\lambda+{\frac {19}{576}}+ \big( -{\frac {13}{
128}}\,\lambda-{\frac {65}{768}} \big) {y}^{2}+ \big( {\frac {21}{
128}}\,\lambda+{\frac {67}{512}} \big) {y}^{4}-{\frac {25}{512}}\,{y
}^{6}
$&$0
$&$ \big( \frac{5}{8}\,{\lambda}^{2}+{\frac {311}{512}}\,\lambda+{\frac {
269}{36864}} \big){y}^{2} + \big( {\frac {185}{768}}\,\lambda+{\frac {3205
}{18432}} \big) {y}^{4}-{\frac {35}{512}}\,{y}^{6}
$&$0
$&$\big( {\frac {455}{4608}}\,\lambda+{\frac {4459}{110592}}
 \big){y}^{4}  -{\frac {21}{1024}}\,{y}^{6}
$&---&---&---\\
5 & $0
$&$\big( {\frac {25}{128}}\,{\lambda}^{3}-{\frac {355}{1536}}\,{
\lambda}^{2}-{\frac {20005}{36864}}\,\lambda-{\frac {11081}{73728}}
 \big)y + \big( {\frac {5}{32}}\,{\lambda}^{2}+{\frac {155}{2048}}\,
\lambda-{\frac {9091}{147456}} \big) {y}^{3}+ \big( {\frac {465}{
2048}}\,\lambda+{\frac {2855}{16384}} \big) {y}^{5}-{\frac {875}{
12288}}\,{y}^{7}
$&$0
$&$\big( {\frac {15}{256}}\,{\lambda}^{2}+{\frac {25}{73728}}\,
\lambda-{\frac {100015}{1769472}} \big){y}^{3} + \big( {\frac {4795}{36864
}}\,\lambda+{\frac {76223}{884736}} \big) {y}^{5}-{\frac {315}{8192}
}\,{y}^{7}
$&$0
$&$\big( {\frac {805}{18432}}\,\lambda+{\frac {26579}{1769472}}
 \big){y}^{5} -{\frac {77}{8192}}\,{y}^{7}
$&---&---\\
6 & ${\frac {25}{32}}\,{\lambda}^{4}+{\frac {95435}{49152}}\,{\lambda}^{3}+
{\frac {862135}{589824}}\,{\lambda}^{2}+{\frac {7959115}{28311552}}\,
\lambda-{\frac {551045}{28311552}}+ \big( {\frac {25}{512}}\,{\lambda
}^{3}-{\frac {355}{6144}}\,{\lambda}^{2}-{\frac {20005}{147456}}\,
\lambda-{\frac {11081}{294912}} \big) {y}^{2}+ \big( {\frac {15}{
256}}\,{\lambda}^{2}+{\frac {465}{16384}}\,\lambda-{\frac {9091}{
393216}} \big) {y}^{4}+ \big( {\frac {775}{8192}}\,\lambda+{\frac {
14275}{196608}} \big) {y}^{6}-{\frac {6125}{196608}}\,{y}^{8}
$&$0
$&$\big( {\frac {335}{256}}\,{\lambda}^{3}+{\frac {9895}{4608}}
\,{\lambda}^{2}+{\frac {440125}{589824}}\,\lambda-{\frac {1755973}{
42467328}} \big){y}^{2} + \big( {\frac {245}{3072}}\,{\lambda}^{2}+{\frac 
{9325}{294912}}\,\lambda-{\frac {31315}{786432}} \big) {y}^{4}+
 \big( {\frac {455}{3072}}\,\lambda+{\frac {32011}{294912}} \big) {
y}^{6}-{\frac {1575}{32768}}\,{y}^{8}
$&$0
$&$\big( {\frac {125}{6912}}\,{\lambda}^{2}-{\frac {2765}{
1769472}}\,\lambda-{\frac {1824697}{79626240}} \big){y}^{4} + \big( {
\frac {10241}{147456}}\,\lambda+{\frac {1504909}{35389440}} \big) {y
}^{6}-{\frac {693}{32768}}\,{y}^{8}
$&$0
$&$\big( {\frac {17549}{884736}}\,\lambda+{\frac {243479}{
42467328}} \big){y}^{6} -{\frac {143}{32768}}\,{y}^{8}
$&---\\
7 & $0
$&$\big( {\frac {25}{64}}\,{\lambda}^{4}-{\frac {40645}{98304}}\,{
\lambda}^{3}-{\frac {2042065}{1179648}}\,{\lambda}^{2}-{\frac {
58120985}{56623104}}\,\lambda-{\frac {39130979}{339738624}} \big)y +
 \big( {\frac {745}{2048}}\,{\lambda}^{3}+{\frac {36385}{73728}}\,{
\lambda}^{2}+{\frac {200065}{2359296}}\,\lambda-{\frac {6542965}{
169869312}} \big) {y}^{3}+ \big( {\frac {645}{8192}}\,{\lambda}^{2}
+{\frac {27925}{786432}}\,\lambda-{\frac {645475}{18874368}} \big) {
y}^{5}+ \big( {\frac {85225}{589824}}\,\lambda+{\frac {1539545}{
14155776}} \big) {y}^{7}-{\frac {25725}{524288}}\,{y}^{9}
$&$0
$&$\big( {\frac {1485}{8192}}\,{\lambda}^{3}+{\frac {439675}{
1769472}}\,{\lambda}^{2}-{\frac {246977}{28311552}}\,\lambda-{\frac {
513938083}{10192158720}} \big){y}^{3} + \big( {\frac {17435}{442368}}\,{
\lambda}^{2}+{\frac {190295}{14155776}}\,\lambda-{\frac {117973177}{
5096079360}} \big) {y}^{5}+ \big( {\frac {36449}{393216}}\,\lambda+
{\frac {6114493}{94371840}} \big) {y}^{7}-{\frac {8085}{262144}}\,{y
}^{9}
$&$0
$&$\big( {\frac {12385}{1769472}}\,{\lambda}^{2}+{\frac {35497}{
28311552}}\,\lambda-{\frac {181782779}{20384317440}} \big){y}^{5} + \big( 
{\frac {5425}{147456}}\,\lambda+{\frac {2961899}{141557760}} \big) {
y}^{7}-{\frac {3003}{262144}}\,{y}^{9}
$&$0
$&$\big( {\frac {32417}{3538944}}\,\lambda+{\frac {375383}{
169869312}} \big){y}^{7} -{\frac {2145}{1048576}}\,{y}^{9}
$\\\end{tabular}

  \end{table}
 \end{landscape}

 \begin{landscape}
 \begin{table}
 \renewcommand{\arraystretch}{1.8}\renewcommand{\baselinestretch}{1.4}
 \caption{Coefficients $\alpha_{li}$ up to the order $q=8$.\label{tab_alpha}}
  \begin{tabular}{c|*{ 2}{>{\footnotesize\PBS\raggedright}p{0.3cm}}
                  *{ 1}{>{\footnotesize\PBS\raggedright}p{1.2cm}}
                  *{ 1}{>{\footnotesize\PBS\raggedright}p{1.7cm}}
                  *{ 1}{>{\footnotesize\PBS\raggedright}p{2.0cm}}
                  *{ 1}{>{\footnotesize\PBS\raggedright}p{2.6cm}}
                  *{ 1}{>{\footnotesize\PBS\raggedright}p{3.0cm}}
                  *{ 1}{>{\footnotesize\PBS\raggedright}p{3.5cm}}
                  *{ 1}{>{\footnotesize\PBS\raggedright}p{4.5cm}}}
{\Large $ _l\!\!\diagdown\!\!^i$} & 0 & 1 & 2 & 3 & 4 & 5 & 6 & 7 & 8  \\  \hline 
1 & $1
$&$0
$&$0
$&$0
$&${\frac {25}{128}}\,{\lambda}^{2}+{\frac {175}{768}}\,\lambda+{\frac {
1225}{18432}}
$&$0
$&${\frac {25}{64}}\,{\lambda}^{3}+{\frac {68075}{98304}}\,{\lambda}^{2}+
{\frac {410275}{1179648}}\,\lambda+{\frac {2568475}{56623104}}
$&$0
$&${\frac {134925}{131072}}\,{\lambda}^{4}+{\frac {239525}{98304}}\,{
\lambda}^{3}+{\frac {316929175}{169869312}}\,{\lambda}^{2}+{\frac {
1001961695}{2038431744}}\,\lambda+{\frac {2116102111}{97844723712}}
$\\2 & ---&$-\frac{1}{8}
$&$0
$&$-{\frac {25}{32}}\,\lambda-{\frac {175}{384}}
$&$0
$&$-{\frac {475}{512}}\,{\lambda}^{2}-{\frac {555}{512}}\,\lambda-{\frac 
{965}{4608}}
$&$0
$&$-{\frac {10025}{4096}}\,{\lambda}^{3}-{\frac {1698785}{393216}}\,{
\lambda}^{2}-{\frac {28823195}{14155776}}\,\lambda-{\frac {123632507}{
679477248}}
$&$0
$\\3 & ---&---&${\frac {5}{16}}\,\lambda+{\frac {17}{96}}
$&$0
$&${\frac {5}{16}}\,{\lambda}^{2}+{\frac {1315}{3072}}\,\lambda+{\frac {
9235}{73728}}
$&$0
$&${\frac {1765}{2048}}\,{\lambda}^{3}+{\frac {1985}{1024}}\,{\lambda}^{2
}+{\frac {4356505}{3538944}}\,\lambda+{\frac {18119345}{84934656}}
$&$0
$&${\frac {2195}{1024}}\,{\lambda}^{4}+{\frac {3052455}{524288}}\,{
\lambda}^{3}+{\frac {888829145}{169869312}}\,{\lambda}^{2}+{\frac {
515947105}{301989888}}\,\lambda+{\frac {13524934807}{97844723712}}
$\\4 & ---&---&---&$-{\frac {5}{256}}\,\lambda-{\frac {107}{6144}}
$&$0
$&$-{\frac {475}{1536}}\,{\lambda}^{2}-{\frac {323095}{884736}}\,\lambda-
{\frac {2199527}{21233664}}
$&$0
$&$-{\frac {15815}{24576}}\,{\lambda}^{3}-{\frac {6138925}{5308416}}\,{
\lambda}^{2}-{\frac {71128549}{113246208}}\,\lambda-{\frac {
12275264429}{122305904640}}
$&$0
$\\5 & ---&---&---&---&${\frac {25}{768}}\,{\lambda}^{2}+{\frac {515}{13824}}\,\lambda+{\frac 
{13777}{1327104}}
$&$0
$&${\frac {25}{384}}\,{\lambda}^{3}+{\frac {163145}{1327104}}\,{\lambda}^
{2}+{\frac {62041}{786432}}\,\lambda+{\frac {133449991}{7644119040}}
$&$0
$&${\frac {20225}{98304}}\,{\lambda}^{4}+{\frac {81963095}{127401984}}\,{
\lambda}^{3}+{\frac {260962627}{382205952}}\,{\lambda}^{2}+{\frac {
10908318845}{36691771392}}\,\lambda+{\frac {1946363428441}{
44030125670400}}
$\\6 & ---&---&---&---&---&$-{\frac {25}{24576}}\,{\lambda}^{2}-{\frac {3749}{1769472}}\,\lambda-{
\frac {384013}{424673280}}
$&$0
$&$-{\frac {425}{12288}}\,{\lambda}^{3}-{\frac {5290135}{84934656}}\,{
\lambda}^{2}-{\frac {82443283}{2264924160}}\,\lambda-{\frac {
6684148507}{978447237120}}
$&$0
$\\7 & ---&---&---&---&---&---&${\frac {125}{73728}}\,{\lambda}^{3}+{\frac {20765}{7077888}}\,{\lambda
}^{2}+{\frac {77707}{47185920}}\,\lambda+{\frac {6177187}{20384317440}
}
$&$0
$&${\frac {125}{24576}}\,{\lambda}^{4}+{\frac {515945}{42467328}}\,{
\lambda}^{3}+{\frac {1166591837}{101921587200}}\,{\lambda}^{2}+{\frac 
{62067170651}{12230590464000}}\,\lambda+{\frac {3643667826661}{
4109478395904000}}
$\\8 & ---&---&---&---&---&---&---&$-{\frac {25}{1179648}}\,{\lambda}^{3}-{\frac {595}{6291456}}\,{\lambda
}^{2}-{\frac {619081}{6794772480}}\,\lambda-{\frac {28211731}{
1141521776640}}
$&$0
$\\9 & ---&---&---&---&---&---&---&---&${\frac {125}{2359296}}\,{\lambda}^{4}+{\frac {41825}{339738624}}\,{
\lambda}^{3}+{\frac {1276387}{12230590464}}\,{\lambda}^{2}+{\frac {
692812541}{17978967982080}}\,\lambda+{\frac {18307106611}{
3451961852559360}}
$\\\end{tabular}

  \end{table}
 \end{landscape}

\label{lastpage}

\end{document}